\documentclass[aps,prb,twocolumn,superscriptaddress,floatfix]{revtex4}
\usepackage{graphicx}
\usepackage{dcolumn}
\usepackage{bm}
\usepackage{color}%
\usepackage{amsmath}
\usepackage{amsfonts}

\begin{document}

\title{Molecular adsorption study of nicotine and caffeine on the 
single-walled carbon nanotube from first principles}

\author{Hyung-June Lee}
\author{Young-Kyun Kwon}
\address{Department of Physics and Research Institute for Basic Sciences, Kyung Hee University, Seoul 130-701, Korea}
\author{Gunn Kim}
\address{Department of Physics and Graphene Research Institute, Sejong University, Seoul 143-747, Korea}

\date{\today}

\begin{abstract}
Using first-principles calculations, we investigate the electronic structures and binding properties of nicotine and caffeine adsorbed
on single-walled carbon nanotubes to determine whether CNTs are appropriate for filtering or sensing nicotine and caffeine molecules.
We find that caffeine adsorbs more strongly than nicotine.
The different binding characteristics are discussed by analyzing the modification of the electronic structure of the molecule-adsorbed CNTs.
We also calculate the quantum conductance of the CNTs in the presence of nicotine or caffeine adsorbates
and demonstrate that the influence of caffeine is stronger than nicotine on the conductance of the host CNT.
\end{abstract}


\maketitle

\section{Introduction}

Carbon nanotubes (CNTs) have attracted much academic and
industrial interest because of their remarkable physical and chemical
properties.~\cite{Chico,Tan, Kwon,Yao,Sanvito,Berber,Zhou,JLee,GKim,
Charlier,Choi} One of the most intriguing characteristics of CNTs is
a very large surface area~\cite{Long1,Cinke,Yin} comparable to that of
activated carbon. This property has been exploited to develop various
applications including hydrogen storage, filters removing toxic
compounds, and chemical sensors. In spite of the overestimated earlier
studies,~\cite{Dillon,Chambers,Chen} however, CNTs did not appear to be
a good storage medium for hydrogen due to their inert surfaces, where
H$_2$ molecules adsorb very weakly via van der Waals
interaction.\cite{Liu,Ahn,Panella,Kwon2} The binding properties of
hazardous molecules on CNTs were also examined theoretically and
experimentally~\cite{Long2,Peng,Fegan1,Fegan2,Tournus} for potential
filter application. One of the most promising applications of CNTs is a
sensor application. CNT-based sensors were developed and demonstrated
to detect a wide range of molecules such as ammonia,~\cite{Kong} nitrogen dioxide,~\cite{Kong} oxygen,~\cite{Collins}
alcohol,~\cite{Song} and other molecules.~\cite{Bradley,Yu,Burt}
They verified that CNT-based sensors are able to detect not only the type of gas molecule, but also the
concentration of the adsorbate.

Nicotine and caffeine are both alkaloids and therapeutic compounds.
They serve as stimulants for the central nervous system in humans.
Through tobacco and coffee, nicotine and caffeine are
addictive drugs used heavily by humans. Because of their psychoactive
effects, nicotine and caffeine could have positive effects. For
example, nicotine appears to enhance concentration and memory due to
the increase of acetylcholine,~\cite{Rusted} and caffeine seems to
enhance performance in endurance sports.~\cite{Bishop,Conger} However,
these drugs often produce side effects. Both of them may increase the
heart rate that further limits the ability of the body to maintain
homeostasis during exercise. Overdose of these drugs causes severe problems.
In particular, nicotine overdose can be deadly, and excessive ingestion of
caffeine over extended periods of time results in a toxic condition
(caffeinism), with symptoms of vomiting, elevated blood pressure, rapid
breathing, heart palpitations, and insomnia.
As beverages containing a lot of caffeine to provide mental or physical stimulation, energy drinks have been recently produced.
Because they contain more caffeine than strong coffee,
some countries have restrictions on the sale and/or manufacture of energy drinks.
Therefore, the fabrication of sensors and filters with high sorption capability of these chemicals
is desired for public health.

In this paper, we report our study on the adsorption properties of nicotine and
caffeine on CNTs. We examined the modification of the electronic
structures of CNTs due to adsorbates. We found that both nicotine and
caffeine have non-covalent interaction with the CNT through $\pi$-stacking.
Although both of nicotine and caffeine contain aromatic rings and nitrogen atoms,
it was found that they show very different binding geometries and binding energies.

\section{Computational details}

We performed first-principles calculations based on the density
functional theory~\cite{Hohenberg,Kohn} to study the structural and
electronic properties of the adsorbates on CNTs. We employed a
plane-wave basis set and the projector augmented wave (PAW) implemented
in the Vienna ab initio Simulation Package (VASP).~\cite{Kresse1,
Kresse2} The cutoff energy for the kinetic energy was set to 500 eV.
The generalized gradient approximation (GGA)~\cite{Perdew}
was used to describe the exchange-correlation energy functional.
In general, chemical bonding and electron transfer
can be described well within the GGA.
However, the dispersion or van der Waals (vdW) forces are not represented well
by the GGA functional because quantum
electronic interactions in the regions of low electron densities are
not correctly expressed. In these calculations, thus we made the correction
of vdW interaction between the adsorbates and the CNT,
using Grimme's method\cite{Grimme} for the GGA (i.e., the GGA plus vdW).

To verify the validity of our computational results, we also repeated
our work with the linear combination of pseudoatomic orbitals
implemented in the SIESTA code.~\cite{SIESTA1,SIESTA2} We used 210 Ry
as the mesh cutoff energy and double-$\zeta$ basis with polarization
for the basis set. All the model structures were relaxed until none of
the residual forces acting on any atom would exceed
0.03~eV/$a_{\rm B}$, where $a_{\rm B}$ is the Bohr radius. The binding
energies were corrected using the basis set superposition error
correction with ghost atoms.
The binding energies, electronic band structures, and wavefunctions presented in this paper were all obtained
using the pseudoatomic orbital basis set implemented in the SIESTA package.
The general trends in the binding energies and the electronic structures for nicotine and caffeine are very similar for the two computational packages.

We selected the semiconducting (8, 0) CNT as the host material for molecular adsorption.
We used a tetragonal supercell with a length of 16.94~\AA~along the CNT axis and 20~\AA~along the lateral
directions. We utilized $1\times1\times4$ Monkhorst-Pack
sampling~\cite{Monkhorst} for the Brillouin zone integrations. On the
other hand, a cubic supercell with a length of 20~\AA~along all three
directions and only one $k$ point at the $\Gamma$ point were used for
calculations of the isolated molecules (nicotine and caffeine). Charge transfer between the adsorbate and the
host CNT was estimated using the Bader charge analysis.
For each configuration, we obtained its binding energy
($E_{\rm b}$) defined by
\[ 
    E_{\rm b}=E[\mbox{CNT}]+E[\mbox{molecule}]-E[\mbox{CNT+molecule}],
\]
where $E[\mbox{CNT+molecule}]$ and $E[\mbox{CNT}]$ are the total
energies of the CNT with and without a nicotine (or caffeine) molecule, respectively,
and $E$[molecule] represents the energy of an isolated nicotine (or caffeine) molecule.

The quantum transport calculations were performed using the TranSIESTA code,
based on density functional theory (DFT), and the self-consistent electronic structure
of a nanostructure connected to the electrodes.
Non-equilibrium Green's functions were used to solve an open system from the DFT Hamiltonian.
In the calculations of the quantum transport, $1\times1\times36$ $k$ points were used in the Monkhorst-Pack scheme.\cite{Monkhorst}

\section{Results and Discussion}

\begin{figure}[t]
  \centering
\includegraphics[width=0.8\columnwidth]{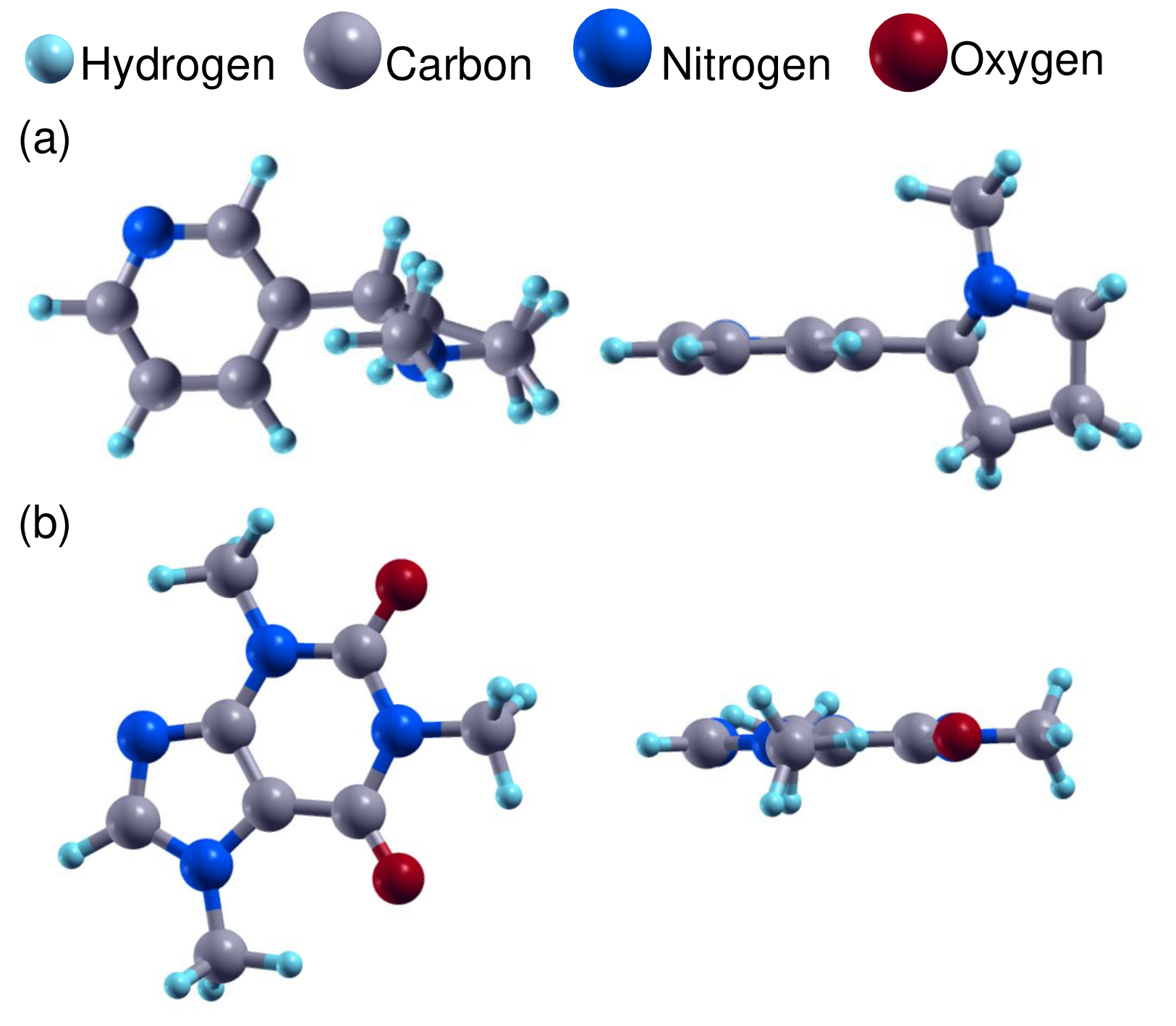}
\caption{Model structures of (a) nicotine and (b) caffeine. Left: top view. Right: side view.}
\label{Fig1}
\end{figure}

First, we calculated the individual nicotine and caffeine
molecules in the vacuum condition to optimize the gas phase geometry.
Figures~\ref{Fig1}a and \ref{Fig1}b show the optimized geometries of nicotine
and caffeine molecules. All the nitrogen atoms form essentially a planar
structure with carbon atoms implying $sp^2$ orbital hybridization,
maintaining the aromatic characteristics.

\begin{figure}[t]
  \centering
\includegraphics[width=0.7\columnwidth]{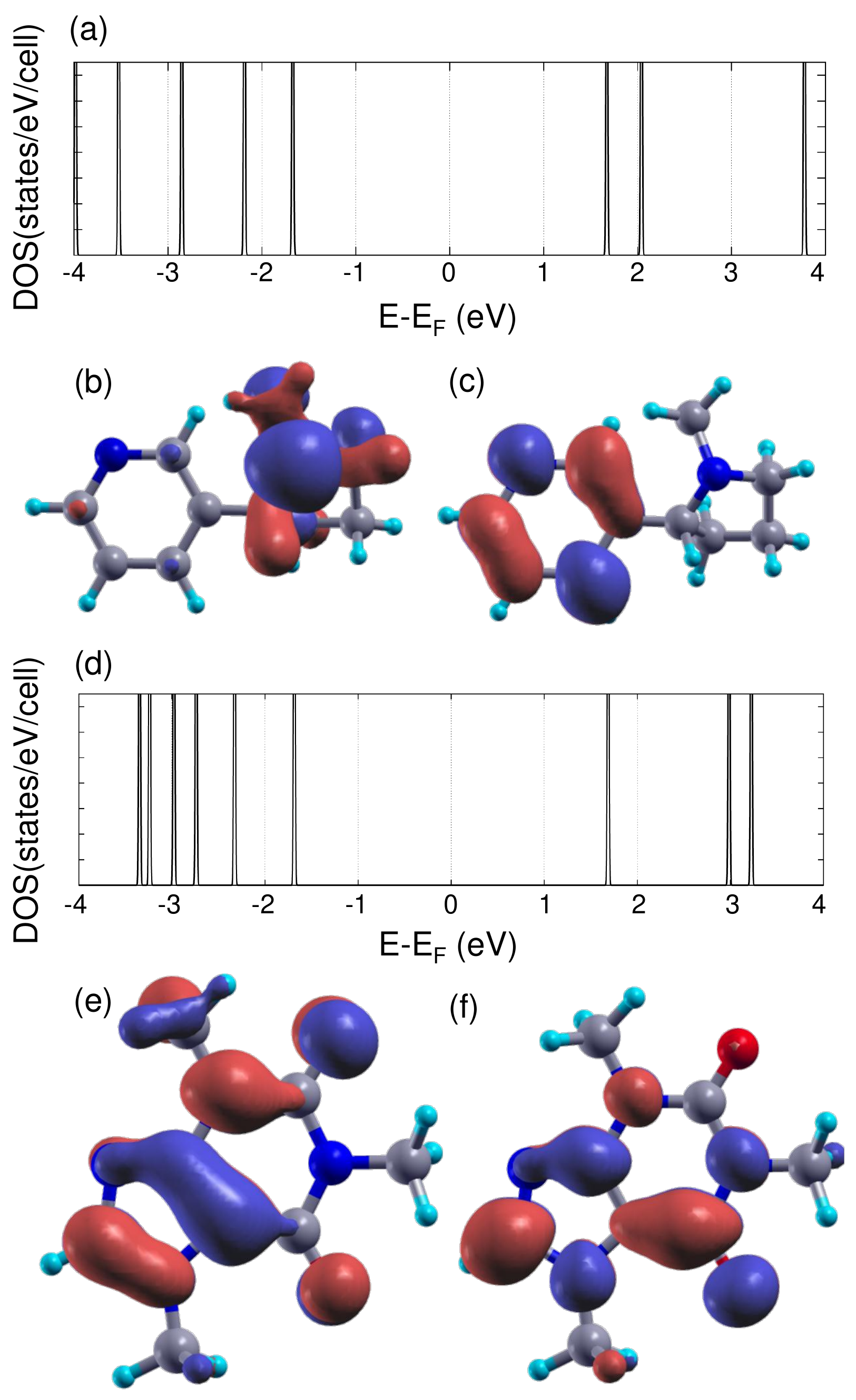}
\caption{(a) DOS, (b) HOMO and (c) LUMO orbitals of a nicotine molecule. (d) DOS, (e) HOMO and (f) LUMO orbitals of caffeine molecule.}
\label{Fig2}
\end{figure}

Figures~\ref{Fig2}a--c show the density of states (DOS), the highest occupied molecular orbital
(HOMO) and the lowest unoccupied molecular orbital (LUMO) of
a nicotine molecule, respectively. Similarly, Figs.~\ref{Fig2}d--f
represent the same quantities of a caffeine molecule. The energy gaps
between the HOMO and the LUMO levels are 3.34~eV and 3.37~eV for the
nicotine and caffeine molecules, respectively, which are very similar
to each other. However, the energy difference between the LUMO and
LUMO+1 (the state one level above the LUMO) of the nicotine molecule
in Fig.~\ref{Fig2}a is much smaller that that of the caffeine
molecule in Fig.~\ref{Fig2}d. Interestingly, the HOMO
of nicotine is dominant at the five-membered ring as shown in Fig.~\ref{Fig2}b,
whereas the electronic density in the LUMO is high at the phenyl group (six-membered ring)
as displayed in Fig.~\ref{Fig2}c. In contrast, the HOMO and LUMO wavefunctions of
caffeine appear to be distributed over the whole molecule as shown in Figs.~\ref{Fig2}e and \ref{Fig2}f.

\begin{figure}[t]
  \centering
\includegraphics[width=0.8\columnwidth]{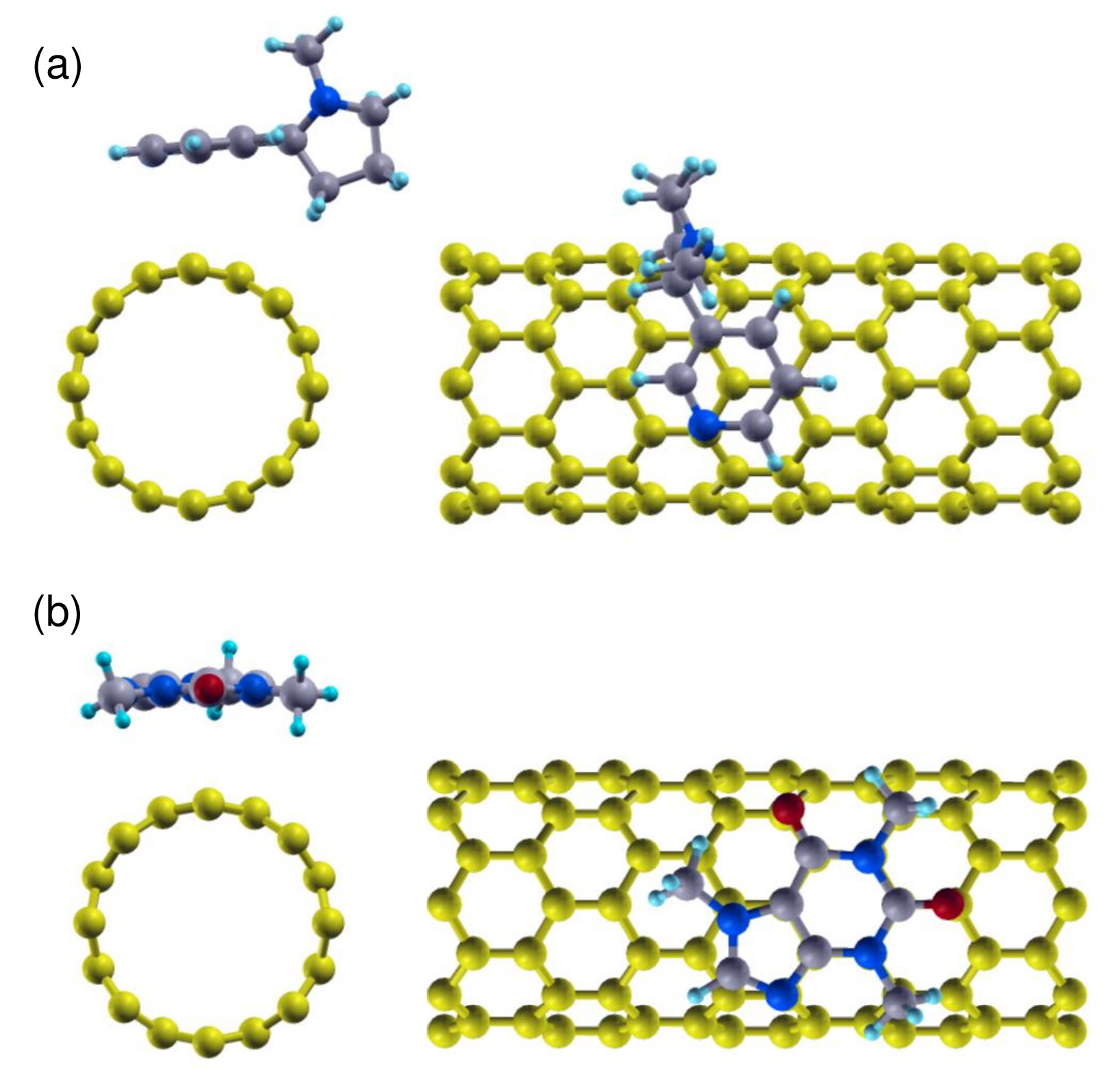}
\caption{Optimized structure of the (8, 0) CNT with (a) an adsorbed nicotine molecule and (b) an adsorbed caffeine molecule. The figures on
the left hand side show side view, whereas the right hand side figures do top view. To avoid confusion, the carbon atoms in the CNT are shown in yellow.}
\label{Fig3}
\end{figure}

Figure~\ref{Fig3} shows the optimized structures of the (8, 0) CNTs with
adsorbed nicotine and caffeine molecules. As shown in Fig.~\ref{Fig3}a,
the nicotine molecule prefers to contact with the CNT
through its six-membered ring rather than its five-membered ring.
Moreover, the most stable binding geometry appears similar to Bernal
(``AB'') stacking of graphite with the binding energy of
$E_b\approx0.46$~eV and the binding distance of $d_b \approx 3.29$~\AA.
On the other hand, the caffeine molecule is bound to the CNT stronger
than the nicotine molecule (almost twice). Its binding energy and
distance are $E_b \approx 0.88$~eV and $d_b \approx 3.25$~\AA,
respectively.
It is noteworthy that according to an {\em ab initio} study of Debbichi {\em et al.}, the binding energy of antracene on the (10, 10) CNT is $\sim$ 1 eV
when the van der Waals interaction is considered.\cite{Debbichi}
Our charge analysis demonstrates that no appreciable charge transfer occurs between the molecules and the CNT.
Then, what makes the difference in the binding energy for the two types of molecules?
Unlike nicotine, a hexagon and a pentagon in caffeine are both on the same plane,
and thus this planar structure enhances the $\pi$ stacking with the CNT surface.
In contrast to AB stacking in the nicotine adsorption,
the caffeine molecule prefers ``AA'' to AB stacking, which is only
$\sim$ 5 meV less stable though.

\begin{figure}[t]
  \centering
\includegraphics[width=0.8\columnwidth]{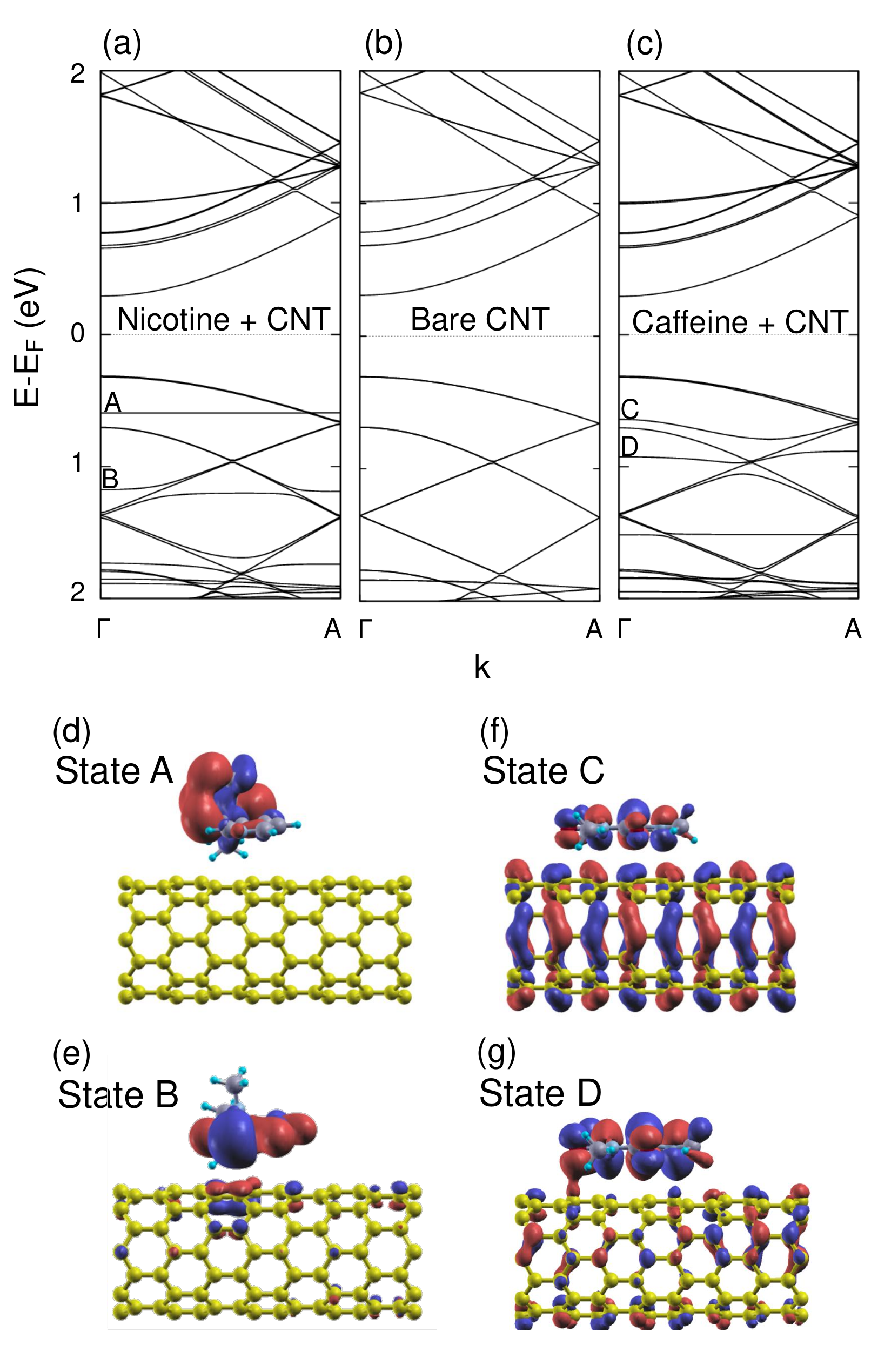}
\caption{The band structures of (a) the nicotine-adsorbed CNT, (b) the bare (8, 0) CNT and (c) the caffeine-adsorbed CNT.
(d-e) The states A and B indicate wavefunctions at labeled ``A'' and ``B'' in (a). (f-g) The states C and D indicate
wavefunctions at labeled ``C'' and ``D'' in (c). The red and blue surfaces indicate the phase of the wave functions.}
\label{Fig4}
\end{figure}

Figure~\ref{Fig4} displays the electronic structures of the bare,
nicotine-, and caffeine-adsorbed (8, 0) CNTs. The bare (8, 0) CNT has an
energy band gap of 0.58 eV (Fig.~\ref{Fig4}b), whereas both nicotine-
and caffeine-adsorbed CNTs increase their band gap to $\sim$ 0.61 eV,
as seen in Figs.~\ref{Fig4}a and \ref{Fig4}c. In Figs.~\ref{Fig4}a--c, we did
not find a significant shift in the Fermi level by the molecular
adsorption implying no significant charge transfer between the
adsorbate and the CNT.  In each of the band structure of the
molecule-adsorbed CNT, there are two flat bands originating from the
molecule. For nicotine (caffeine), such flat bands are located
near 0.6~eV (0.6~eV) and 1.3~eV (0.9~eV) below the Fermi level.
Figures~\ref{Fig4}d and \ref{Fig4}e display the wavefunctions  corresponding
to the states labeled ``A'' and ``B'' in Fig.~\ref{Fig4}a at the
$\Gamma$ point. The state A in Fig.~\ref{Fig4}d shows a localized
state corresponding to the HOMO of nicotine, whereas the
state B exhibits a weakly hybridized $\pi$ state on the CNT together
with the nicotine HOMO$-$1 state (the state just below the HOMO).
As shown in Fig.\ref{Fig2}b, the electronic density in the HOMO of nicotine is mainly at the five-membered ring and thus
it has no practical coupling with any CNT state.
Similarly, we show in Figs.~\ref{Fig4}f and \ref{Fig4}g the wavefunctions at the $\Gamma$
point for the two states labeled ``C'' and ``D'' for the caffeine adsorption in Fig.~\ref{Fig4}c.
Interestingly, due to the stronger interaction of caffeine with the CNT than that of nicotine via $\pi$-stacking, the state C corresponding to the HOMO state of
caffeine shows interaction between an extended state of the CNT and a molecular orbital of caffeine.

In 2010, Gir\H{a}o et al. investigated nicotine adsorption on the single-wall CNT\cite{Girao}.
Thus, we need to compare the differences between their models and ours.
For the binding energy, our model has stronger adsorption than their models for nicotine adsorbed on the CNT.
For the band structure, their models have localized states originating from the nicotine molecule in the forbidden band (just above the valence band top).
In contrast, our most stable model for nicotine-adsorbed CNT has no localized state in the forbidden band (see Fig. \ref{Fig4}).

\begin{figure}[t]
  \centering
\includegraphics[width=1.0\columnwidth]{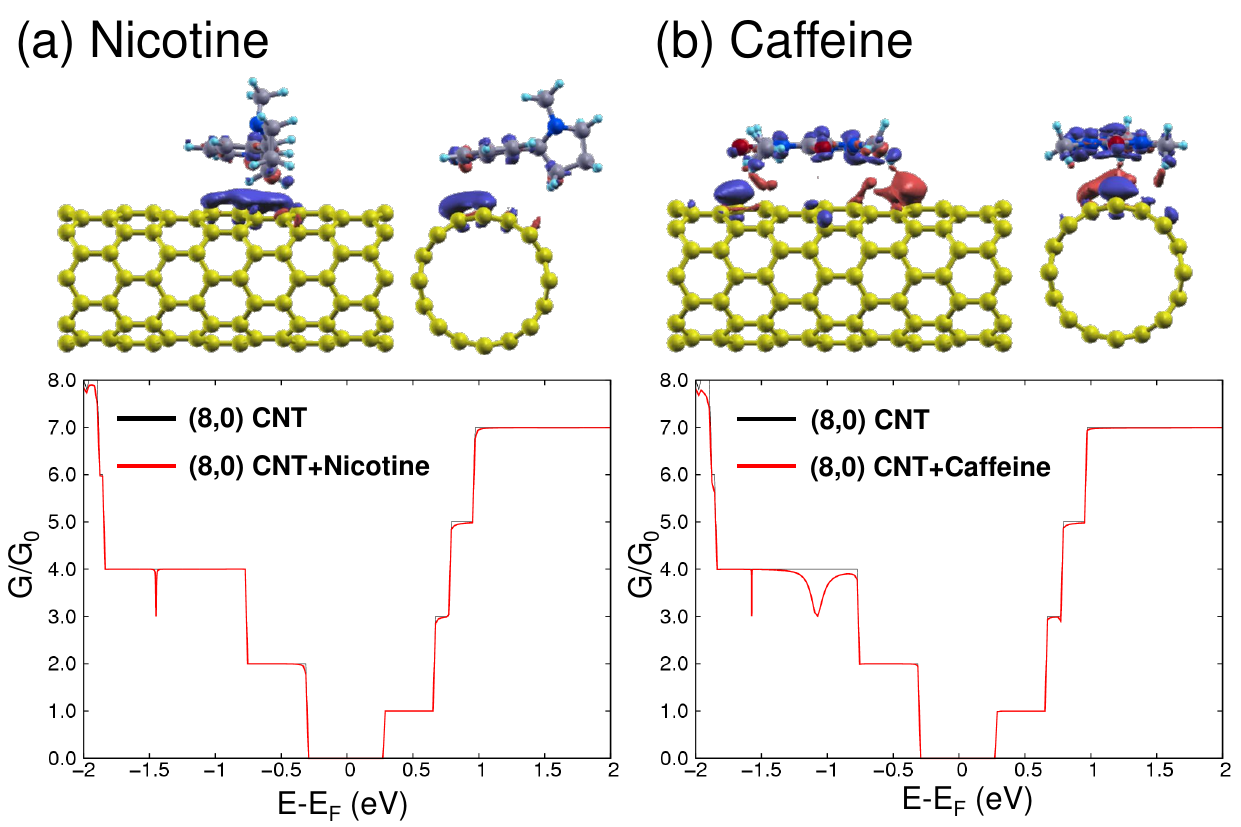}
\caption{Charge density difference and quantum conductance plots of (a) nicotine- and (b) caffeine-adsorbed
CNTs. In the charge density difference plots, red and blue colors represent regions of electron accumulation and depletion, respectively.
Here, $G_{0}$ (= $2e^{2}/h$) represents the conductance quantum, where $e$ is called the elementary charge ($\approx 1.602 \times 10^{-19}$ C) and $h$ is the Planck constant
($\approx 1.436 \times 10^{-15}$ eV$\cdot$s).
}
\label{Fig5}
\end{figure}

For further analysis, we calculated charge density differences shown in
Fig.~\ref{Fig5}. Isodensity surfaces indicate the regions of electron
accumulation (red) and depletion (blue).
It is shown that the nicotine molecule creates a local polarized region at the CNT near the adsorbate as shown
in Fig.~\ref{Fig5}b.
We found a more interesting feature in the caffeine-adsorbed CNT,
where an electron depletion takes place near the O atom bonded to the C atom between two N atoms in the six-membered ring of the caffeine molecule.
At the same time, an electron accumulation region is formed at the opposite end of the adsorbate (near the five-membered ring).
This phenomenon can be interpreted as a ``donation and back-donation'' between the CNT and the caffeine molecule in the axial direction of the CNT.
It is noteworthy that the DFT calculation does not describe accurately the amount of charge transfer but it guarantees the overall shape of isodensity surfaces.

Finally, we present the quantum conductance plots of the (8, 0) CNTs with the adsorbed nicotine and caffeine molecules.
As shown in Fig. \ref{Fig5}, only caffeine gives rise to a broad conductance dip about 1.1 eV below the Fermi level
with the reduction of conductance from 4$G_0$ to 3$G_0$,
which is attributed to the state C in Fig. \ref{Fig4}f. Here, $G_{0}$ (= $2e^{2}/h$) represents the conductance quantum,
where $e$ is called the elementary charge ($\approx 1.602 \times 10^{-19}$ C) and $h$ is the Planck constant
($\approx 1.436 \times 10^{-15}$ eV$\cdot$s).
On the other hand, both of nicotine and caffeine molecules give rise to very narrow dips in conductance (from 4$G_0$ to 3$G_0$) around 1.5 eV below the Fermi level,
which are related to the states B and D in Figs. \ref{Fig4}e and \ref{Fig4}g, respectively.
As mentioned above, the state C corresponds to the HOMO of caffeine strongly coupled to the extended CNT state, whereas
the states B and D originate from the HOMO$-$1 states of nicotine and caffeine and they are very weakly hybridized with CNT states, respectively.
We note that the energy levels of conductance dips may have an offset with those of energy eigenstates.
The wavefunctions and the electronic band structures are obtained for a finite CNT with an adsorbed molecule in periodic boundary conditions.
In contrast, the quantum conductance is calculated for a finite CNT with an adsorbate molecule connected to two semi-infinite perfect CNTs (electrodes) in both sides.
Therefore, the energy offset may occur between the quantum electron conductance and enegy eigenstates.
Consequently, the caffeine molecule using a CNT-based device can be monitored in the current-voltage ({\em I-V}) measurement.
In the case of nicotine, on the other hand, it is difficult to detect the molecule using a CNT-based electronic device.
Therefore, one should use the CNTs to filter nicotine molecules.

\section{Conclusion}

In conclusion, we have investigated the electronic structures and binding
properties of nicotine and caffeine molecules on (8, 0) single-walled
carbon nanotubes using first-principles calculations. We found that
both adsorbates show different binding characteristics on CNTs.
The nicotine molecule binds to the CNT with an adsorption energy of 0.46 eV,
whereas the caffeine molecule adsorbs with a higher binding energy of 0.88 eV.
To figure out the difference, we discussed the modification of the
electronic structures of the nicotine- and caffeine-adsorbed CNTs.
Although there is no significant charge transfer between the CNT and the adsorbate,
we found that there is an noticeable interaction between them. In particular,
the caffeine-adsorbed CNT exhibits a special binding feature
formed by electron donation and back-donation.
Our study of quantum transport demonstrates that {\em I-V} curves can distinguish nicotine from caffeine using a CNT-based electronic sensor.
We believe our findings of the adsorption properties of caffeine and nicotine on the CNT
will help the future studies of intermolecular interaction between the host nanotube and other organic molecules.

\section*{Acknowledgement}
G. K. acknowledges the Priority Research Center Program (Grant No. 2010-0020207) and
the Basic Science Research Program through the National Research Foundation of Korea (NRF) (Grant No. 2013R1A1A2009131).
Y. K. acknowledges the financial support from the National Research Foundation of Korea (Grant No. 2011-0016188).
Some portion of our computational work was done using the resources of the KISTI Supercomputing Center (KSC-2012-C2-19 and KSC-2012-C2-72).

\end{document}